\documentclass[10pt,letterpaper]{article}
\usepackage{opex3}

\usepackage{soul}

\usepackage{adjustbox}
\usepackage{color}
\usepackage{graphicx}
\usepackage{subfigure}
\usepackage{amsmath}
\usepackage{citesort}
\usepackage{bm}

\newcommand{\comment}[1]{}

\newcommand{\ie}{\mbox{i.e.\ }}
\newcommand{\eg}{\mbox{e.g.\ }}

\newcommand{\eqnref}[1]{Eq.~(\ref{#1})}
\newcommand{\tabref}[1]{Table~\ref{#1}}
\newcommand{\figref}[1]{Fig.~\ref{#1}}

\newcommand{\eps}{\varepsilon}

\newcommand{\Power}{{\mathcal{P}}}
\newcommand{\Length}{\Lambda}
\newcommand{\total}{{\rm{d}}}

\newcommand{\PowerGain}{\Gamma}
\newcommand{\TotalGain}{\mathcal{A}}
\newcommand{\FOM}{\mathcal{F}}
\newcommand{\Ge}{\text{Ge}}
\newcommand{\Si}{\text{Si}}

\renewcommand{\vec}[1]{{\bf{#1}}}

\newcommand{\mode}[2]{{#1}^{(#2)}}

\newcommand{\TPA}{{\text{2PA}}}
\newcommand{\FCA}{{\text{FCA}}}

\newcommand{\opt}{{\text{opt}}}

\newcommand{\TLSCONE}{C_1}
\newcommand{\TLSCTWO}{C_2}

\begin{document}

\title{Power limits and a figure of merit for stimulated Brillouin scattering
in the presence of third and fifth order loss}

\author{C. Wolff,$^{1,2\ast}$
P. Gutsche,$^{2,3}$
M.~J. Steel,$^{1,4}$
B.~J. Eggleton,$^{1,5}$
and C.~G. Poulton$^{1,2}$}
\address{
  \begin{centering}
  $^1$ Centre for Ultrahigh bandwidth Devices for Optical Systems (CUDOS), Australia;
  \newline
  $^2$ School of Mathematical and Physical Sciences, \newline
  University of Technology Sydney, NSW 2007, Australia;
  \newline
  $^3$ Zuse Institute Berlin, Takustra{\ss}e 7, D–14195 Berlin, Germany;
  \newline
  $^4$ MQ Photonics Research Centre, Department of Physics and Astronomy, \newline
  Macquarie University Sydney, NSW 2109, Australia;
  \newline
  $^5$ Institute of Photonics and Optical Science (IPOS), School of Physics, \newline
  University of Sydney, NSW 2006, Australia.
  \end{centering}
}

\email{christian.wolff@uts.edu.au}

\date{\today}

\begin{abstract}
  We derive a set of design guidelines and a figure of merit to aid the 
  engineering process of on-chip waveguides for strong Stimulated Brillouin 
  Scattering (SBS).
  To this end, we examine the impact of several types of loss on the total 
  amplification of the Stokes wave that can be achieved via SBS.
  We account for linear loss and nonlinear loss of third 
  order (two-photon absorption, 2PA) and fifth order, most notably 2PA-induced 
  free carrier absorption (FCA).
  From this, we derive an upper bound for the output power of continuous-wave 
  Brillouin-lasers and show that the optimal operating conditions and maximal 
  realisable Stokes amplification of any given waveguide structure are 
  determined by a dimensionless parameter $\FOM$ involving the SBS-gain and 
  all loss parameters.
  We provide simple expressions for optimal pump power, waveguide length and
  realisable amplification and demonstrate their utility in two example systems.
  Notably, we find that 2PA-induced FCA is a serious limitation to SBS in 
  silicon and germanium for wavelengths shorter than $2200\,\text{nm}$ and 
  $3600\,\text{nm}$, respectively.
  In contrast, three-photon absorption is of no practical significance.
\end{abstract}


\ocis{() ;
      () ;
      () .}

\section{Introduction}
Stimulated Brillouin Scattering (SBS) is the coherent and self-amplifying 
interaction between light waves and a hypersonic acoustic wave that are confined 
in the same waveguide or bulk material~\cite{Boyd2003}.
Initially predicted by Brillouin~\cite{Brillouin1922} and first observed in
quartz~\cite{Chiao1964}, it has been well known for many years in optical fibers
as a highly resonant and strong third-order nonlinearity in both
backward~\cite{Agrawal} and forward directions~\cite{Kang2009}.
More recently, there has been an intense research effort to generate SBS in 
on-chip waveguides: 
SBS has been observed in chalcogenide glass rib
waveguides~\cite{Pant2011} and, inspired by theoretical
considerations~\cite{Qiu2013}, silicon nanowires~\cite{Shin2013,vanLaer2015}.
These experiments have greatly broadened the applicability of SBS to 
include a large number of on-chip applications, such as powerful narrow-band light
sources~\cite{Kabakova2013,Buettner2014}, non-reciprocal light
propagation~\cite{Huang2011,Aryanfar2014}, slow light~\cite{Thevenaz2008}
and signal processing in the context of microwave 
photonics~\cite{Vidal2007,Li2013,Morrison2014}.

The realisation of SBS in any device hinges on achieving sufficient Stokes 
amplification within the device length to be useful. 
Within the basic theory, where SBS is the only nonlinear process present, 
the amplification of the Stokes wave (\ie the ratio of Stokes power at the 
output to injected Stokes power) is proportional to the power of the injected 
pump beam; the Stokes wave initially
exhibits exponential growth until it starts to deplete the pump.
In reality, however, neither this exponential growth, nor the predicted linear
relationship between amplification and pump power, can be expected over the 
whole range of pump powers. 
The reason for this is nonlinear loss in the waveguide, where third-order 
processes -- particularly two-photon absorption (2PA) -- impact the 
SBS-performance quite differently from fifth-order processes such as 
three-photon absorption (3PA) and 2PA-induced free carrier absorption (FCA).
The latter is a major nonlinear loss mechanism in group-IV semiconductors such 
as silicon and germanium in important wavelength ranges.
These materials are of interest because many promising applications of SBS 
require the integration of SBS-gain in group-IV photonic circuits in order 
to benefit from CMOS-type fabrication
techniques for mass production and integration with electronic circuitry.

In this paper, we apply recent analytic work~\cite{Wolff2015b} to examine the 
limitations that nonlinear loss imposes on the gain for SBS.
The related effect of nonlinear loss in Stimulated Raman Scattering has been
investigated~\cite{Rukhlenko2010} although based on different approximations
and with an emphasis on short-pulse propagation and pump depletion.
In contrast, we present two important findings: firstly, that there is an upper
bound for the output Stokes \emph{power} that can be obtained with a specific waveguide design 
(\ie for a specific geometry of the waveguide cross-section with a specific choice of materials) by amplifying a weak (\eg thermal) initial Stokes signal. This bound, which
involves both nonlinear loss coefficients and the SBS-gain parameter of the 
design, in particular represents an absolute upper bound for the output power of 
any continuous wave (CW) Brillouin-laser.
Secondly, we find that the total \emph{amplification} (ratio of output to input power) 
of an incident Stokes wave that can be obtained with a specific waveguide 
design also has an upper bound.
This maximally realisable amplification and the corresponding optimal operating
conditions are determined by a dimensionless \emph{figure of merit},
which involves only the effective coefficients for the linear
waveguide loss, the 2PA, the 2PA-induced FCA and the SBS-gain coefficient.
Our figure of merit forms a valuable engineering tool for the design of 
SBS-active waveguides since it provides a convenient way to compare the 
performance of different designs and yields good estimates for the waveguide 
length and pump power.
We demonstrate this in the last section of the paper using two types of
suspended silicon nanowires in backward SBS configuration as examples.

\section{Approach}

We choose to describe the problem within our coupled-mode 
framework~\cite{Wolff2015a} for SBS in an integrated waveguide, which is 
assumed to extend in the positive $z$-direction.
When focusing on strongly confining waveguides, as in this paper, 
it is appropriate to express the optical fields in terms of
power carried by the individual waveguide modes, although other formulations are
possible, \eg based on field intensities as is common in bulk nonlinear 
optics.
The equations of motion for the power $\mode{P}{1}(z)$ carried in the Stokes 
wave and the power $\mode{P}{2}(z)$ in the pump wave are 
\begin{align}
  s \partial_z \mode{P}{1} &=  
  (\PowerGain - 2 \beta - \gamma \mode{P}{2}) \mode{P}{2} \mode{P}{1}
  -\alpha \mode{P}{1} 
  + \TLSCONE, 
  \label{eqn:ss_stokes}
  \\
  \partial_z \mode{P}{2} &=   
  -(\beta + \gamma \mode{P}{2}) [\mode{P}{2}]^2
  -\alpha \mode{P}{2} 
  + \TLSCTWO,
  \label{eqn:ss_pump}
\end{align}
where the symbol $s$ is used to describe both forward SBS ($s=+1$) and 
backward SBS ($s=-1$) within a common framework.
Here $\PowerGain = \int \total r^2 \vec u^\ast \cdot \vec f$ is the 
overlap integral between the optical force distribution
(involving both electrostriction and radiation pressure) as known from the 
literature~\cite{Wolff2015a,Qiu2013}.
The acoustic angular frequency $\Omega$ is defined by the acoustic
dispersion relation and the optical wave number via the usual phase
matching condition~\cite{Wolff2015a,Qiu2013}.
Later we will also provide theoretical estimates for the Brillouin line width
$\Delta \Omega$, which we compute from the overlap integral~\cite{Wolff2015a}
between the acoustic eigenmodes and the dynamic viscosity of silicon and 
germanium~\cite{Helme1978}.

The linear loss coefficient $\alpha$, 2PA-coefficient $\beta$, and 
coefficient  $\gamma$ for 2PA-induced FCA are overlap integrals of the
optical eigenmodes and the respective dissipative nonlinear polarization
currents as presented \eg in Ref.~\cite{Wolff2015b}.

The two large-signal correction terms
\begin{align}
  \TLSCONE = & 
  -(\beta + \gamma \mode{P}{1} + 4 \gamma \mode{P}{2}) [\mode{P}{1}]^2,
  \label{eqn:lsc_stokes}
  \\
  \TLSCTWO = & 
  -(2\beta + \PowerGain + 4 \gamma \mode{P}{2} + \gamma \mode{P}{1})
  \mode{P}{1} \mode{P}{2}
  \label{eqn:lsc_pump}
\end{align}
can be neglected when assuming that the Stokes wave is weak compared to the
pump wave, which we refer to as the small-signal approximation.
Within this approximation, forward and backward SBS yield the same results
for the power and amplification limits.
We have assumed quasi-CW light and have absorbed the acoustic envelope
into $\PowerGain$ via a local acoustic response approximation. This is
justified if the optical modes vary on a length scale large compared to
the acoustic decay length~\cite{Wolff2015a}; in particular, the optical loss
must be sufficiently low~\cite{vanLaer2015b}.

Equations~(\ref{eqn:ss_stokes}--\ref{eqn:lsc_pump}) are derived in 
detail in Ref.~\cite{Wolff2015b} and we restrict ourselves here to a brief
description of the nonlinear loss terms.
The terms involving $\beta$ describe 2PA due to power carried in the respective
mode itself (prefactor $\beta$) and due to the combination of both modes
(prefactor $2\beta$).
The nontrivial factor 2 arises because interference between the modes leads to
a non-uniform intensity distribution; it appears commonly in nonlinear 
optics~\cite{Rukhlenko2010}.
The terms involving $\gamma$ describe fifth-order loss (mainly 2PA-induced
FCA) and as before appear as loss induced by either mode alone (prefactor 
$\gamma$) and cross terms that describe the contribution of inter-mode
interference to the 2PA-induced FCA (prefactor $4 \gamma$).
The nontrivial factor 4 arises in a similar fashion~\cite{Wolff2015b} 
as in the case of 2PA, but is less commonly encountered.

For waveguides with eigenmodes that differ strongly from plane waves 
(\eg silicon nanowires), 
the coefficients $\Gamma, \alpha, \beta$ and $\gamma$ depend sensitively
on the geometry and eigenmode field distributions in the transversal plane.
In the case of waveguides with plane wave-like eigenmodes (\eg silica fibers), 
the optical powers can be replaced with local intensities and the parameters 
with established parameters~\cite{Boyd2003,Agrawal}.
This is equivalent to normalising the eigenmode (\ie plane wave) over a cross 
section of $1\,\text{m}^2$ within our framework.

Finally, we note that inter-mode SBS can be described by the same equations at 
the expense of a more convoluted notation, leading to analogous results with 
modified parameters (see appendix).

\section{Maximal output power}
We begin with the output power upper bound.
From the large-signal Stokes equation 
[\ie Eqs.~(\ref{eqn:ss_stokes},\ref{eqn:lsc_stokes})] 
we find
\begin{align}
  s \partial_z \mode{P}{1} < [\PowerGain - 2 \beta - 4 \gamma \mode{P}{1} 
  - \gamma \mode{P}{2}] \mode{P}{1} \mode{P}{2}.
  \label{eqn:power_est_start}
\end{align}
In conjunction with the requirement $s \partial_z \mode{P}{1} > 0$ expressing 
Stokes amplification, this provides a necessary 
(though not tight)
upper bound for the output power:
\begin{align}
  \mode{P}{1} < \frac{\PowerGain - 2 \beta - 4 \gamma \mode{P}{2}}{4 \gamma} < \frac{\Power}{4}
  \ ,
  \label{eqn:power_limit}
\end{align}
where we have introduced the natural unit of power 
\begin{align}
  \Power = (\PowerGain - 2 \beta) / \gamma.
  \label{eqn:def_power}
\end{align}
Any greater Stokes power level will necessarily decay with propagation
and the maximum power limit can only be exceeded anywhere in the waveguide
if it is exceeded at the input face.
Notably, \eqnref{eqn:power_limit} is an absolute upper bound for the 
output power of any CW Brillouin-laser.
This should be distinguished from the influence of linear loss, which mainly 
influences the laser threshold and the line width.
Note also that the simple expression $\Power/4$ overestimates the maximal 
output power because of the term $4\gamma\mode{P}{2}$ and further nonlinear
loss terms, which were removed in arriving at the inequality 
\eqnref{eqn:power_est_start}.

\section{Maximal amplification and optimal design parameters}

\begin{figure}
  \centering
  \includegraphics[width=0.65\columnwidth]{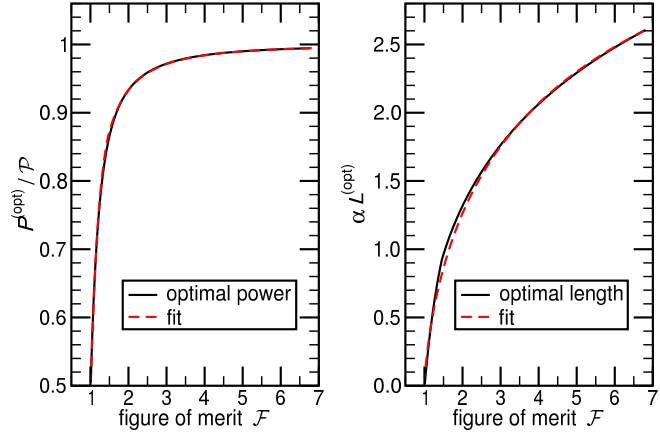}

  \caption{
    Optimal pump power (left panel) and optimal waveguide length
    (right panel) as functions of the figure of merit $\FOM$ 
    [see \eqnref{eqn:fom}];
    numerical results (black solid lines) in comparison to the approximate 
    expressions \eqnref{eqn:pump_approx} and \eqnref{eqn:length_approx} 
    (red dashed lines)
    The optimal pump power is between $\Power / 2$ and $\Power$ 
    and the optimal waveguide length is of order $\alpha^{-1}$.
  }
  \label{fig:opt_parm}
\end{figure}

The discussion of optimal design parameters and figures of merit is based on 
the small-signal equations, \ie Eqs.~(\ref{eqn:ss_stokes},\ref{eqn:ss_pump}) 
under the approximation 
\mbox{$\TLSCONE = \TLSCTWO = 0$}. 
As a result, we obtain upper bounds for both the pump wave and the Stokes wave 
power levels, and, therefore, for the measurable total Stokes amplification 
of a given waveguide.
This can be concluded from the fact that the neglected terms 
Eqs.~(\ref{eqn:lsc_stokes},\ref{eqn:lsc_pump}) are strictly negative.
The small-signal equations can be solved analytically for the case $\beta = 0$
and starting from this solution, the case of $\beta \neq 0$ can be treated 
perturbatively~\cite{Wolff2015b}.
In order to strip the problem of as many free parameters as possible, it is
advisable to express all powers with respect to the natural power unit $\Power$ 
introduced in \eqnref{eqn:def_power} and to express all lengths with respect to 
the corresponding natural length unit 
\begin{align}
  \Length = \gamma / (\PowerGain - 2 \beta)^2.
  \label{eqn:def_length}
\end{align}
We thus obtain an analytical expression that predicts the total Stokes
amplification $\TotalGain$ as a function of the waveguide length $L$ and the 
injected pump power $\mode{P}{2}(0) = P_0$:
\begin{align}
  \TotalGain(L, P_0) = & 10 \log_{10} \left[\frac{\mode{P}{1}(L)}{\mode{P}{1}(0)}\right] \ \text{dB}
  \\
  \nonumber
  = & 
  \frac{10}{\ln 10} \bigg\{
    \frac{1}{\sqrt{\alpha \Length}} \left[
      \tan^{-1} \sqrt{(1 + \psi) \exp(2 \alpha L) - 1} - \tan^{-1} \psi
    \right]
    \\
    & \quad - \frac{1}{2} \ln \left[ \frac{(1 + \psi)\exp(2\alpha L) - 1}{\psi} \right]
  \bigg\} \ \text{dB},
  \label{eqn:amplification}
\end{align}
where $\psi = \alpha / (\gamma P_0^2)$ is a measure of the relative strength
of linear loss and 2PA-induced FCA.
The factor 
$1/\sqrt{\alpha\Length} = (\PowerGain - 2 \beta)/\sqrt{\alpha\gamma}$ 
in front of the first term of \eqnref{eqn:amplification} is of particular
interest, because it arises naturally from \eqnref{eqn:ss_stokes} by 
requiring that a positive $\mode{P}{2}$ exists such that 
\mbox{$s \partial_z \mode{P}{1} > 0$} (\ie Stokes amplification at some 
positive pump power).
Since all coefficients are positive, this is only possible if 
\begin{align}
  \mode{P}{2} = \frac{1}{2\gamma} \left[ (\PowerGain - 2\beta) \pm \sqrt{(2\beta - \Gamma)^2 - 4\alpha \gamma} \right] > 0;
  \quad \Rightarrow \quad \frac{\PowerGain - 2 \beta}{2 \sqrt{\alpha \gamma}} > & 1 \ .
\end{align}
This is a necessary condition for any amplification of the Stokes wave.
Therefore the quantity 
\begin{align}
  \FOM = \frac{\PowerGain - 2 \beta}{2 \sqrt{\alpha \gamma}} 
  \label{eqn:fom}
\end{align}
lends itself as a \emph{figure of merit} for SBS in
waveguide designs.

\begin{figure}
  \centering
  \includegraphics[width=0.65\columnwidth]{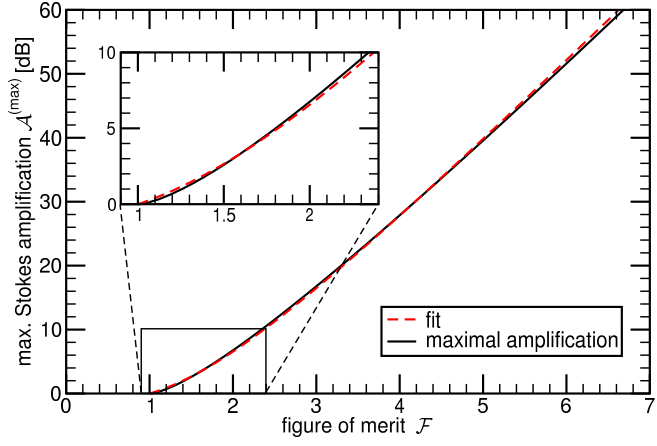}

  \caption{
    The maximally realisable Stokes amplification for FCA-dominated 
    setups as a function of the figure of merit $\FOM$
    [see \eqnref{eqn:fom}] computed numerically (black solid line)
    and approximate fit according to \eqnref{eqn:ampl_approx}
    (red dashed line).
    Materials and waveguide designs with $\FOM < 1$ cannot amplify an
    injected Stokes wave.
  }
  \label{fig:amplification}
\end{figure}

Every $\FOM$ corresponds to a different maximally realisable Stokes 
amplification $ \TotalGain^{(\text{max})} $ and, through~\eqref{eqn:amplification}, 
this amplification is only
obtained for a specific (optimal) choice of waveguide length $L^{(\opt)}$ 
and pump power $P^{(\opt)}$.
We determined these values by numerically finding the maximum of
$\TotalGain(L, P_0)$ for a wide range of $\FOM$ and show them as 
black solid lines in  \figref{fig:opt_parm} and 
\figref{fig:amplification}.
Observe that the worse the figure of merit, the weaker the pump and shorter 
the waveguide that delivers the optimal performance: since FCA is a 
fifth-order process while the SBS gain and 2PA are third-order processes, 
trying to drive a poor system harder is counterproductive.

\begin{table*}[t!]
  \caption{
    Table of nonlinear coefficients ($\PowerGain$, $\beta$, $\gamma$),
    acoustic frequency $\Omega / 2\pi$, Brillouin line width 
    $\Delta \Omega / 2\pi$, SBS-figure of merit $\FOM$,
    maximally realisable Stokes amplification $\TotalGain^{\text(max)}$ 
    and the natural unit of power $\Power$ (four times the max. SBS-laser 
    output power in case of $\FOM > 1$) for \emph{bulk} silicon and germanium at
    wavelengths on the red and the blue side of the 2PA-threshold.
    We assumed a linear loss of $\alpha = 0.1\,\text{dB/cm}$ and a carrier
    life time of $10\,\text{ns}$ throughout.
    The quantities $\PowerGain$ and $\Omega$ were computed from literature 
    expressions~\cite{Wolff2015a} assuming wave propagation along the 
    [100]-direction and purely electrostrictive coupling based on literature 
    photoelastic coefficients~\cite{Biegelsen1974,Feldman1978}.
    The Brillouin line width was computed as described in the main text using
    literature values for the dynamic viscosity~\cite{Helme1978}.
    The bulk 2PA-, 3PA- and FCA-coefficients are literature
    values~\cite{Bristow2007,Seo2011,Pearl2008} (annotated in the square brackets).
    As mentioned in the main text, this table lists bulk parameters,
    which are naturally expressed as power \emph{intensities} 
    (units of W/m${}^2$) rather than powers; this is reflected in the units
    for $\TotalGain$, $\beta$, $\gamma$ and $\Power$ and $P^{(\text{opt})}$,
    which differ from those in \tabref{tab:waveguides}.
    Note that the photo-elastic tensor and the dynamic viscosity are
    available in the literature only for selected wavelengths, so we assumed
    them to be nondispersive, leading to a wavelength-independent prediction
    of the SBS-gain.
  }
  \label{tab:bulk}

  \medskip

  \begin{tabular}{l||c|c|c||c|c|c||c}
    & \multicolumn{3}{c||}{Silicon} &  \multicolumn{3}{c||}{Germanium} & Unit \\ \hline
    $\lambda_0$ & $1550$ & $2000$ & $2400$ & $3000$ & $3600$ & $4000$ & nm
    \\
    \hline
    \hline
    $\Omega${\footnotesize $ / 2\pi$} & $38.1$ & $29.5$ & $24.6$ & $13.2$ & $11.0$ & $9.90$ & GHz
    \\
    $\Delta \Omega$ {\footnotesize $ / 2\pi$} & $162$ & $97$ & $30$ & $39$ & $27$ & $22$ & MHz
    \\
    $\PowerGain$ & $2.55$ & $2.55$ & $2.55$ & $324$ & $324$ & $324$ & mm/GW
    \\
    \hline
    $\alpha$ & 2.3 & 2.3 & 2.3 & 2.3 & 2.3 & 2.3 & m${}^{-1}$
    \\
    $\beta$ & $15\,{}^{\text{\cite{Bristow2007}}}$ & $3\,{}^{\text{\cite{Bristow2007}}}$ & $0^{\ast}$ & 
    $200\,{}^{\text{\cite{Seo2011}}}$ & $0.9\,{}^{\text{\cite{Seo2011}}}$ & $0^{\ast}$ & mm/GW
    \\
    $\gamma\,{}^\dagger$ & $0.29\,{}^{\text{\cite{Bristow2007}}}$ & $0.08\,{}^{\text{\cite{Bristow2007}}}$ & $3.5${\footnotesize $\times 10^{-5}$}$\,{}^{\text{\cite{Pearl2008}}}$ 
    & $ 240\,{}^{\text{\cite{Seo2011}}}$ & $ 1.1\,{}^{\text{\cite{Seo2011}}}$ & $2.7${\footnotesize $\times 10^{-4}$}$\,{}^{\text{\cite{Seo2011}}}$ & mm/(GW${}^2$)
    \\
    \hline
    \hline
    $\Power$ & -- & -- & $7200$ & -- & $0.29$ & $1200$ & W/($\mu$m${}^2$)
    \\
    $\FOM$ & $-0.54$ & $-0.13$ & $4.5$ & $-0.06$ & $3.2$ & $410$ & --
    \\
    \hline
    $L^{(\opt)} $ & -- & -- & 0.95 & -- & 0.79 & 2.6 & m
    \\
    $P^{(\opt)} $ & -- & -- & 7100 & -- & 0.28 & 1200 & W/($\mu$m${}^2$)
    \\
    $\TotalGain^{(\text{max})}$ & -- & -- & $34$ & -- & $19$ & $\gg 60^\ddagger$ & dB
  \end{tabular}

  \medskip
  {
    \footnotesize
    \begin{tabular}{rl}
      $\ast$ & three photon absorption is the leading multiphoton process at this wavelength
      \\
      $\dagger$ & assuming 10~ns carrier life time
      \\
      $\ddagger$ & only a very rough lower bound since our fit 
      [\eqnref{eqn:ampl_approx}] only covers the range $\FOM < 7$
    \end{tabular}
  }
\end{table*}

The numerically obtained relationships between $\FOM$ (in the range 
$\FOM > 1$) and the performance and operation parameters can be represented 
by three phenomenological fit functions
\begin{align}
  L^{(\opt)} \approx &
  \left( 2 \ln \FOM\right)^{0.713} \alpha^{-1},
  \label{eqn:length_approx}
  \\
  P^{(\opt)} \approx &
  \left( 1 - 0.25 \FOM^{-2} - 0.25 \FOM^{-6} \right) \Power,
  \label{eqn:pump_approx} 
  \\
  \TotalGain^{(\text{max})} \approx &
  13.0 \left[ \sqrt{\FOM (\FOM - 1) + 3.0} - \sqrt{3.0} \right] \ \text{dB}.
  \label{eqn:ampl_approx} 
\end{align}
These expressions are provided as a convenient way to predict the maximal 
amplification and optimal waveguide length and pump power in the process of
designing an SBS-active waveguide.
Their good accuracy over a wide parameter range is demonstrated by 
\figref{fig:opt_parm} and \figref{fig:amplification}, where they are shown 
as red dashed curves.

As mentioned, the interpretation of \eqnref{eqn:fom}
as a figure of merit is
based on the fact that a material or waveguide design with $\FOM<1$ 
cannot amplify a Stokes wave.
For $\FOM>1$, Stokes amplification is possible, but the realisable
amplification factor is limited to a number given by \eqnref{eqn:ampl_approx}.
For values $0 < \FOM < 1$, SBS generation of the Stokes wave occurs in pump-probe experiments 
with externally injected Stokes seeds, but with net attenuation along the
waveguide.
However, the regime of $\FOM > 1$ can still be reached by reducing the linear or
fifth-order loss terms \eg via free carrier extraction techniques.
Finally, for $\FOM < 0$, SBS is overcome by the competing two-photon
absorption irrespective of pump power (since both SBS and 2PA are third-order 
processes) or waveguide length, and no SBS-allowed regime can be reached by 
adjusting $\alpha$ or $\gamma$.

\section{Examples and Conclusions}

\begin{table*}[t]
  \caption{
    Table of SBS-resonance parameters, loss parameters, SBS-figure of merit, 
    maximally realisable Stokes-amplification and optimal operating conditions
    for two simple suspended nanowire designs (see \figref{fig:waveguides}) operated
    in backward SBS configuration in the 2PA-regime.
    The effective waveguide coefficients were computed in analogy to and using
    material parameters listed in \tabref{tab:bulk}, yet using the specified 
    optical and best-suited acoustic eigenmodes.
    The SBS-gain coefficient includes both electrostriction and radiation 
    pressure terms as described in the literature~\cite{Wolff2015a}.
    Furthermore, we show data for an experimentally studied 
    geometry~\cite{vanLaer2015}, where we adopted the SBS-parameters (gain, 
    shift, linewidth), the linear loss and the carrier life time published for
    that particular structure and computed the nonlinear loss and from this 
    $\FOM$, $\Power$ and and extrapolated $\TotalGain^{(\text{max})}$.
    Optimal operation conditions cannot be provided, because this structure does
    not provide net Stokes amplification.
  }
  \label{tab:waveguides}

  \medskip

  \begin{tabular}{l||c|c|c||c|c|c||c||c}
    & \multicolumn{3}{c||}{$E_x$-wire} & \multicolumn{3}{c||}{$E_y$-wire} & Ref.~\cite{vanLaer2015} & \\
    & \multicolumn{3}{c||}{BSBS}       & \multicolumn{3}{c||}{BSBS}       & FSBS  & Unit \\ \hline
    $\lambda_0$ & 1550 & 2000 & 2000 & 1550 & 2000 & 2000 & 1550 & nm
    \\
    width & $315^\ast$ & $406^\ast$ & $406^\ast$ & $372^\ast$ & $480^\ast$ & $480^\ast$ & $450^\ddagger$ & nm
    \\
    height & $284^\ast$ & $365^\ast$ & $365^\ast$ & $256^\ast$ & $330^\ast$ & $330^\ast$ & $230^\ddagger$ & nm
    \\
    \hline
    $\Omega${\footnotesize $ / 2 \pi$} & $12.3$ & $9.56$ & $9.56$ & $10.2$ & $7.92$ & $7.92$ & $9.2^\ddagger$ & GHz
    \\
    $\Delta \Omega${\footnotesize $ / 2 \pi$} & $8.37$ & $5.03$ & $5.03$ & $4.79$ & $2.88$ & $2.88$ & $30^\ddagger$ & MHz
    \\
    $\PowerGain$ & $1210$ & $729$ & $729$ & $4910$ & $2950$ & $2950$ & $3218^\ddagger$ & $\text{W}^{-1}\text{m}^{-1}$
    \\
    \hline
    $\alpha$ & $23$ & $23$ & $4.6$ & $23$ & $23$ & $4.6$ & $60^\ddagger$ & $\text{m}^{-1}$
    \\
    $\beta$ & 380 & 46 & 46 & 290 & 36 & 36  & 84 & $\text{W}^{-1}\text{m}^{-1}$
    \\
    $\gamma\,{}^\dagger$ & 138000 & 13000 & 13000 & 105000 & 9900 & 9900 & 43800 & $\text{W}^{-2}\text{m}^{-1}$
    \\
    $\tau_c$ & $10$ & $10$ & $10$ & $10$ & $10$ & $10$ & $6^\ddagger$ & ns
    \\
    \hline
    \hline
    $\Power$ & 3 & 49 & 49 & 41 & 290 & 290 & 18 & mW
    \\
    $\FOM$ & 0.13 & 0.58 & 1.30 & 1.52 & 1.68 & 5.31 & 0.94 & --
    \\
    \hline
    $L^{(\opt)} $ & -- & -- & 137 & 38 & 45 & 514 & -- & mm
    \\
    $P^{(\opt)} $ & -- & -- & 39 & 36 & 261 & 287 & -- & mW
    \\
    $\TotalGain^{(\text{max})} $ & $< 0$ & $< 0$ & $1.4$ & $2.8$ & $3.9$ & $44$ & $-0.2^\diamond$ & dB
  \end{tabular}

  \medskip
  {
    \footnotesize
    \begin{tabular}{rl}
      $\ast$ & geometry is scaled with the operating wavelength
      \\
      $\dagger$ & computed assuming the carrier life $\tau_c$
      \\
      $\ddagger$ & value taken from Ref.~\cite{vanLaer2015}
      \\
      $\diamond$ & extrapolated using \eqnref{eqn:ampl_approx} 
      slightly outside its intended parameter range
    \end{tabular}
  }
\end{table*}

We demonstrate the utility of our figure of merit by discussing bulk silicon,
bulk germanium, and three (two theoretical, one experimentally realised) silicon 
nanowire systems. 
In the case of bulk materials, the previously presented theory applies directly
on replacing powers with power densities and effective waveguide loss and
gain coefficients with the respective bulk material parameters.
This is presented in \tabref{tab:bulk} for the case of silicon and germanium 
at wavelengths around the 2PA-threshold.
Since this table contains numbers for SBS in bulk, only backward-SBS is possible
and the SBS-gain is entirely defined by electrostriction and photoelasticity; 
radiation pressure does not contribute in \tabref{tab:bulk}.
The SBS-gain and acoustic parameters were computed according to established 
methods for waveguides~\cite{Qiu2013,Wolff2015a} using periodic boundary 
conditions in the transverse plane and literature values for the mechanical
and photoelastic parameters~\cite{Wortman1965,Biegelsen1974,Feldman1978}. 
The acoustic line width $\Delta \Omega$ was computed from the acoustic group
velocity and the acoustic decay parameter $\alpha_{\text{ph}}$, which in turn 
was computed from the dynamic viscosity~\cite{Wolff2015a,Helme1978}.
The nonlinear loss parameters are literature 
values~\cite{Bristow2007,Seo2011,Pearl2008}.
It is clearly visible that bulk silicon and germanium are ill-suited for SBS
on the blue side (shorter wavelengths, higher frequencies) of their respective 
2PA-thresholds ($\approx 2300\,\text{nm}$ and $\approx 3650\,\text{nm}$, 
respectively). 
This is mainly due to the combination of insufficient photoelastic coupling,
high acoustic frequency (entailing high mechanical loss) and the strong 
fifth-order absorption term caused by 2PA-induced free carriers.
However, we find that the fifth-order loss caused by 3PA allows for substantial
figures of merit and should be negligible in practice.
Therefore, SBS in bulk and quasi-bulk silicon and germanium seems infeasible
in the near-IR and should be studied in the mid-IR.

\begin{figure}[t]
  \centering
  \includegraphics[width=0.80\columnwidth]{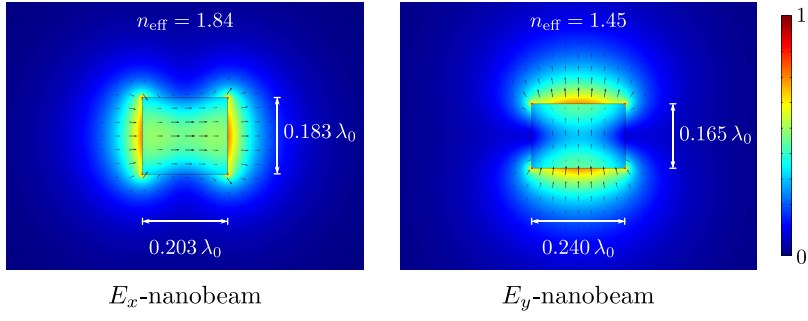}
  \caption{
    Waveguide geometries and effective indices of the optical modes that are
    studied for backward SBS in \tabref{tab:waveguides}.
    The color-plot depicts the modulus of the modal electric field in arbitrary
    units, the black arrows indicate the in-plane electric field components.
    The waveguides consist of silicon in [100]-orientation; their 
    cross-sections are shown as a thin black rectangle and are scaled with the 
    light's vacuum wavelength $\lambda_0$.
  }
  \label{fig:waveguides}
\end{figure}

Furthermore, we present numerical calculations of figures of merit for two 
theoretical nanowire designs at two different wavelengths and one experimental 
silicon nanowire at $1550\,\text{nm}$.
The two theoretical structures together with the relevant optical modes are 
depicted in \figref{fig:waveguides}; note that the second example is not 
operated with the fundamental optical mode.
For both waveguides, we computed the backward-SBS coupling (involving both 
electrostriction and radiation pressure) with the lowest few acoustic modes 
and present the results for the lowest quasi-longitudinal mode.
Again, we computed the SBS-parameters and effective nonlinear coefficients 
using established methods~\cite{Qiu2013,Wolff2015a,Wolff2015b} and literature 
material parameters~\cite{Wortman1965,Biegelsen1974,Helme1978,Bristow2007}.
The results are shown in \tabref{tab:waveguides}.
For the theoretical structures, we have to choose a value for the
linear loss, a quantity which is still improving with fabrication developments.
We present results for $1\,\text{dB/cm}$ (as a typical current value)
for calculations at both wavelengths (columns 2,3,5,6),
 and for $0.2\,\text{dB/cm}$ as a state of the art value for the calculations
at $\lambda_0=$2000~nm only (columns 4,7).

The results indicate that both designs can support SBS (\ie $\FOM > 0$) at both considered 
wavelengths.
However, neither of them is expected to provide net Stokes amplification 
on the blue side of the 2PA-threshold unless the linear loss is reduced 
considerably below $1\,\text{dB/cm}$.
We find it noteworthy that our proposal for a waveguide based on the 
$E_y$-polarized eigenmode is predicted to provide higher SBS-gain than the 
previously studied $E_x$-polarized eigenmode.

For the comparison with the experimental structure, we adopted the geometry,
published forward SBS-parameters and linear loss of a nearly suspended 
silicon nanowire~\cite{vanLaer2015} and complemented this with our numerically
computed nonlinear loss coefficients based on the carrier life time measured
in that structure.
The first observation is that we find a figure of merit just below $1$ 
corresponding to an extrapolated total Stokes amplification of 
$-0.2\,\text{dB}$ in agreement with the published value of $-0.1\,\text{dB}$.
As the natural power unit we find $\Power = 18\,\text{mW}$, which is 
$1.5\,\text{dB}$ less than the published 2PA-threshold of $25\,\text{mW}$;
however, care should
be taken when comparing the results, since the experimental 2PA-threshold gain 
was determined from Fig.~3a in Ref.~\cite{vanLaer2015}, which depicts an 
on/off-gain, while we derived $\Power$ in the context of optimal net Stokes
amplification.

In their paper, the authors of the experimental study mention that their 
device is close to net amplification.
In fact, a reduction of the linear loss from the reported $2.6\,\text{dB/cm}$
to $1\,\text{dB/cm}$ would lead to a $\FOM = 1.5$, which would allow for a 
maximum net Stokes amplification of $2.7\,\text{dB}$ realised for a waveguide 
of $37\,\text{mm}$ and a pump power of $16\,\text{mW}$.
In contrast, a reduction of the carrier life time to $2\,\text{ns}$ would lead
to a similar $\FOM = 1.6$, but the maximum Stokes amplification of 
$3.4\,\text{dB}$ would be achieved for a length of only $16\,\text{mm}$ and a
considerably higher pump power of $48\,\text{mW}$.
The output powers of hypothetical CW SBS-laser using those two improved 
structures would be limited to less than $9\,\text{mW}$ and $27\,\text{mW}$, 
respectively.
This example illustrates how the results and parameter fits presented in this
paper can help in designing the optimal length of samples and hint at the best
route to improve performance.

In summary, we have studied the impact of third-order and fifth-order loss on 
the process of SBS in semiconductor waveguides.
First, we derived an upper bound to the output power of an SBS-based amplifier.
Second, we introduced a figure of merit as a simple measure of the suitability
of an SBS-design and provide simple fit functions that predict the 
SBS-performance of that structure as well as the optimal operating conditions.

We acknowledge financial support from the Australian Research Council (ARC) 
via the Discovery Grant DP130100832, its Laureate Fellowship 
(Prof. Eggleton, FL120100029) program and the ARC Center of Excellence CUDOS 
(CE110001018).

\begin{appendix}

\section{Appendix: Expressions for inter-mode coupling}

Within the main text, we restrict ourselves to intra-mode coupling with the 
goal of cleaner notation.
In this appendix, we provide the corresponding results for the general case 
albeit at the expense of a great number of subscript indices.
The governing equations (see Ref.~\cite{Wolff2015b}) for the general case of 
inter-mode coupling are:
\begin{align}
  s \partial_z \mode{P}{1} &=  
  \underbrace{
    (\PowerGain - 2 \beta_{12} - \gamma_{122} \mode{P}{2}) \mode{P}{2} \mode{P}{1}
    -\alpha_1 \mode{P}{1} 
  }_{\text{small signal terms}}
  -
  \underbrace{
    (\beta_{11} + \gamma_{112} \mode{P}{1} + 4 \gamma_{111} \mode{P}{2}) [\mode{P}{1}]^2
  }_{\text{large signal corrections}} \ ,
  \label{eqn:stokes}
  \\
  \partial_z \mode{P}{2} &=   
  \underbrace{
    -(\beta_{22} + \gamma_{222} \mode{P}{2}) [\mode{P}{2}]^2 
    -\alpha_2 \mode{P}{2} 
  }_{\text{small signal terms}}
  -
  \underbrace{
    (2\beta_{21} + \PowerGain + 4 \gamma_{221} \mode{P}{2} + \gamma_{211} \mode{P}{1})
    \mode{P}{1} \mode{P}{2}
  }_{\text{large signal corrections}} \ .
  \label{eqn:pump}
\end{align}
Note that in contrast to intra-mode coupling, where the same loss coefficient 
($\alpha$, $\beta$, $\gamma$) appears in several terms, these symmetries are now
lifted because the coefficient in each loss term is associated with a different
combination of eigenmodes.
These are the power-related coefficients required in 
Eqs.~(\ref{eqn:stokes},\ref{eqn:pump}) and have been derived from the 
amplitude-related coefficients derived in the Appendix of Ref.~[18]:
\begin{align}
  \alpha_i = &\frac{2 \eps_0 \omega}{\mode{\Power}{i}} \int \total^2 r \ 
  \big| \mode{\widetilde{\vec e}}{i} \big|^2 \Im\{\eps_r\} \ ,
  \label{eqn:appx_linear}
  \\
  \beta_{ij} = & 
  \frac{2}{\mode{\Power}{i} \mode{\Power}{j}} \int \total^2 r \ \big(
  \big| \mode{\widetilde{\vec e}}{i} \cdot \mode{\widetilde{\vec e}}{j} \big|^2 + 
  \big| \mode{\widetilde{\vec e}}{i} \cdot (\mode{\widetilde{\vec e}}{j})^\ast \big|^2 + 
  \big| \mode{\widetilde{\vec e}}{i} \big|^2 \big| \mode{\widetilde{\vec e}}{j} \big|^2 \big)
  \Sigma^\TPA \ ,
  \label{eqn:appx_tpa}
  \\
  \gamma_{ijk} = & 
  \frac{2}{\mode{\Power}{i} \mode{\Power}{j} \mode{\Power}{k}}
  \int \total^2 r \ |\mode{\widetilde{\vec e}}{i}|^2
  \Big[ |\mode{\widetilde{\vec e}}{j} \cdot \mode{\widetilde{\vec e}}{k}|^2 + 
    |\mode{\widetilde{\vec e}}{j} \cdot (\mode{\widetilde{\vec e}}{k})^\ast|^2 +
    |\mode{\widetilde{\vec e}}{j}|^2 |\mode{\widetilde{\vec e}}{k}|^2 
  \Big]
  \Sigma^\FCA \ .
  \label{eqn:appx_fca}
\end{align}
Here, $\mode{\vec e}{1/2}$ represents the eigenmode patterns, 
$\mode{\Power}{1/2}$ the corresponding power flux of the normalised eigenmode
and $\Sigma^\TPA$ and $\Sigma^\FCA$ are the nonlinear conductivities used to
describe two-photon absorption (2PA) and 2PA-induced free carrier absorption
(see Ref.~\cite{Wolff2015b}).

The Stokes power limit follows from the full Eq.~(\ref{eqn:stokes}) exactly in
the same way as described in the main text:
\begin{align}
  \mode{P}{1} < \frac{\PowerGain - 2 \beta_{12}}{4 \gamma_{112}} \ .
  \label{eqn:limit}
\end{align}

The figure of merit follows from the small signal terms of Eq.~(\ref{eqn:stokes})
alone, again exactly as described in the main text for the case of intra-mode
coupling [note that the $\gamma$-coefficient appearing in the denominator is 
not necessarily the same as the one in Eq.~(\ref{eqn:limit})]:
\begin{align}
  \FOM = \frac{\PowerGain - 2 \beta_{12}}{2 \sqrt{\alpha_1 \gamma_{122}}} \ .
\end{align}

\end{appendix}

\end{document}